\documentclass[11pt]{article}
\usepackage{paspconf}
\usepackage{epsfig}
\usepackage{amssymb}
\begin{document}

\title{The role of convective boundaries}
\author{F.\ Herwig}
\affil{Institut f\"ur Physik, Universit\"at Potsdam, Germany} 
\author{T.\ Bl\"ocker}
\affil{Max-Planck-Institut f\"ur Radioastronomie, Bonn, Germany}
\author{D.\ Sch\"onberner}
\affil{Astrophysikalisches Institut Potsdam, Germany}

\begin{abstract}
We investigate the influence of convective overshoot
on stellar evolution models of the thermal pulse AGB phase
with $M_\mathrm{ZAMS}=3\mathrm{M}_\odot$. 
An exponential diffusive overshoot 
algorithm 
is applied to all convective boundaries during all evolutionary
stages. 

We  demonstrate that overshooting at the bottom of the
pulse-driven convective zone, which forms in the intershell during the He-shell flash,
leads to more efficient third dredge-up. Some overshoot
at the bottom of the convective envelope removes the
He-H discontinuity, which would otherwise prohibit the occurrence
of the third dredge-up for this stellar mass. However, no correlation
between the amount of envelope overshoot and the efficiency of the
third dredge-up has been found.

Increasingly efficient third dredge-up eventually leads
to a carbon star model.  
Due to the partial mixing efficiency in the overshoot region a 
\ensuremath{^{13}\textrm{C}}-pocket can form after the third dredge-up event which may be crucial for
$n$-capture nucleosynthesis.          

\end{abstract}


\section{Introduction}
Current stellar evolutionary models of the AGB stage fail to
reproduce a number of observationally based constraints. 
The observed carbon star luminosity function 
        requires that the third dredge-up in AGB stars occurs
        efficiently at core masses of about $\ensuremath{M_{\rm H}}=0.6\mathrm{M}_\odot$ 
        (Groenewegen et~al\ 1995, Marigo et~al\ 1996).

For the explanation of \mbox{$s$-process} elements on the surface
        of AGB stars, dredge-up is an important prerequisite as well.
        Obviously, dredge-up enables the transport of processed
        material to the surface. But dredge-up is also needed 
        in order to establish a region in the star
        where the H-rich convective envelope has immediate
        contact with the \ensuremath{^{12}\textrm{C}}-rich intershell region.
Additionally to sufficient dredge-up a mechanism must
        be identified to ingest protons into the intershell
        region in order to produce the neutron source
        \ensuremath{^{13}\textrm{C}} via the reaction chain $\ensuremath{^{12}\textrm{C}}(\textrm{p},\gamma)\ensuremath{^{13}\textrm{N}}(e^+\nu)\ensuremath{^{13}\textrm{C}}$
        (Gallino et~al\ 1998).

A certain class of post-AGB stars
        (about $20\%$ of all central stars of planetary nebulae)
        show a hydrogen -- deficient surface composition. The 
        evolutionary origin of the typical
        abundance pattern of these stars [e.g.\ (He/C/O)=(33/50/17) in 
        mass fractions for the PG1159 stars, Werner et~al\ (1998)]  
        are not well understood. One possible evolutionary
        scenario establishes a link between
        these surface abundances with the intershell abundance of
        the AGB stars
        (Iben \& McDonald 1995).
        Current models, however, typically predict an abundance
        pattern like (\ensuremath{^{4}\textrm{He}}/\ensuremath{^{12}\textrm{C}}/\ensuremath{^{16}\textrm{O}})=(70/26/1) for the intershell region.

In Sec.\,\ref{sec:overhoot} we will describe the motivation and 
method of overshooting which we have applied to stellar evolution 
models of AGB stars. 
In Sec.\,\ref{sec:dup-pdcz} we focus on the effect of overshooting of the
pulse-driven convective zone, which evolves in the intershell during the He-flash, on the
intershell abundance. We will also
 discuss the role of overshooting at the bottom of the envelope
convection zone for the third dredge-up. 
This will then lead to the presentation of the \ensuremath{^{13}\textrm{C}}-pocket which also 
results from the overshooting treatment (Sec.\,\ref{sec:cdr-pocket}).

\section{Overshooting with decreasing mixing efficiency}
\label{sec:overhoot}

Overshooting is the mixing of material beyond the boundary
of convection. The impact of overshooting has been studied 
for different stellar conditions in order to address various
problems in stellar evolution models.
For example, Shaviv \& Salpeter (1973)
and Maeder (1975) have studied the general properties 
of overshooting at the boundary of a convective core.
The effect of overshooting at the bottom of the convective envelope
has been studied by Alongi et~al\ (1991) for red giant branch 
stars and by Iben (1976) for AGB stars.

Mixing due to overshoot has often been treated
by the formal extension of the instantaneously mixed convective region
(Schaller et~al 1992).
However, hydrodynamic simulations (Freytag et~al 1996) of the
surface convection zone of A-stars and white dwarfs show
prominent convective rolls and downdrafts which extend beyond the
boundary of convection and lead to a partial mixing of this region. 
The exponential decay of the turbulent velocity 
field and the diffusive character of the associated mixing are 
common features of this overshoot. These results have motivated the
ansatz for an exponentially declining diffusion coefficient for the
overshoot region in order to describe the mixing beyond the convective
boundary (Herwig et~al 1997). Further on, we assume that the 
exponential diffusive description of overshoot is applicable  to the deep 
stellar interior. For this region we have scaled the exponential
coefficient $f$ by fitting the main sequence 
width. We then assumed that the
resulting exponential coefficient of $f=0.016$ can serve as an
approximation for the convective boundaries encountered in AGB stellar models.

This method of \emph{exponential diffusive overshoot} has been
applied to all convective boundaries during the AGB evolution (and
before). This includes the pulse driven convective zone in the 
intershell layer which results from the huge energy release during
the He-shell flash. 

\section{The dredge-up in the pulse-driven convective zone and the third dredge-up}
\label{sec:dup-pdcz}
\begin{figure}
\plotone{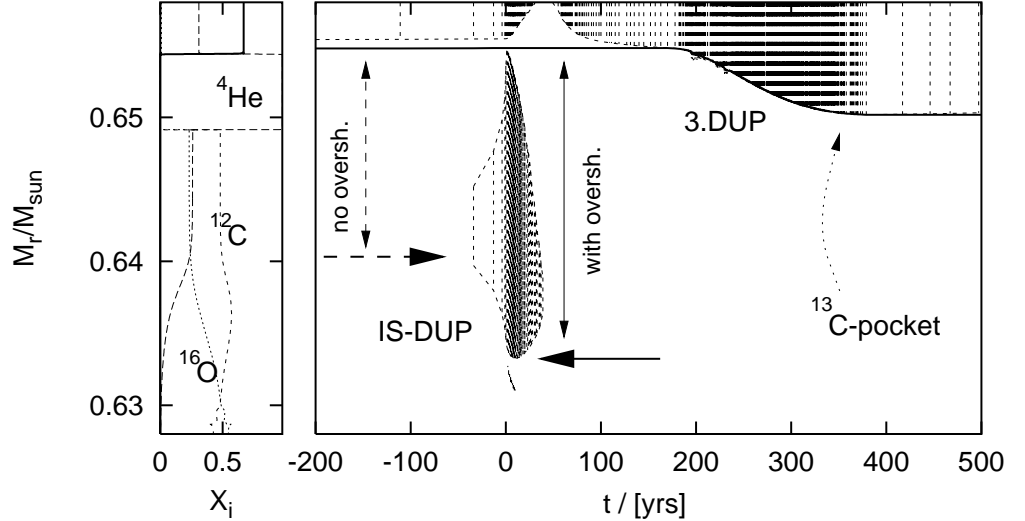}
\caption{\label{fig:KD} 
Isotopic abundance profiles of \ensuremath{^{4}\textrm{He}}, \ensuremath{^{12}\textrm{C}} and
\ensuremath{^{16}\textrm{O}} at the end of the
interpulse phase (left) and  the evolution of convective regions during a 
thermal pulse (at $t=0 \mathrm{yr}$) including the third dredge-up (main panel)
at the 9$^\mathrm{th}$ TP of a
$M_\mathrm{ZAMS}=3\mathrm{M}_\odot$ sequence.
In the main panel the regions covered with vertical dashed lines 
are convectively unstable (convective envelope at the top and the
pulse-driven convective zone in the center of the figure). The density of these lines indicate
the variation of the time resolution.
The dashed arrows (labeled \textsf{no oversh.}) indicate the
extend that a  pulse-driven convective zone during a TP with this core mass
would have without application of overshoot
to this convective region. Label \textsf{IS-DUP}:\emph{intershell dredge-up}.
}
\end{figure}
\begin{figure}
\plotone{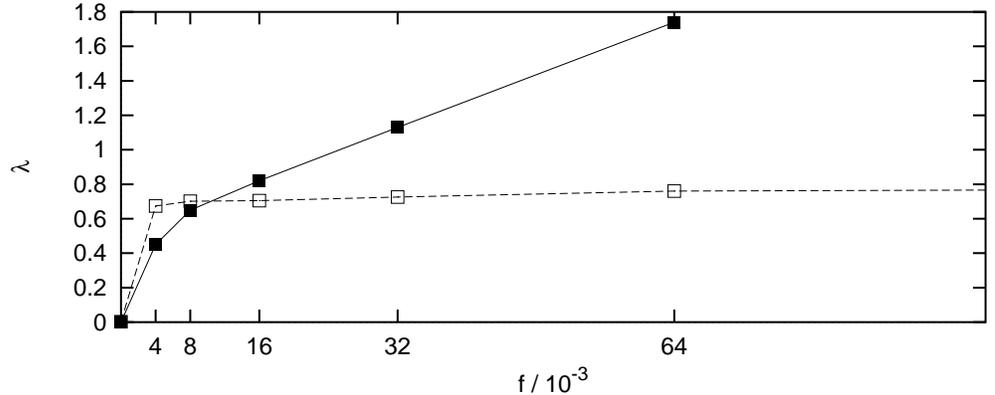}
\caption{ \label{fig:f-lambda} 
The variation of $\lambda$ [($\lambda$ =  (core mass decrease by dredge-up) / 
(core mass growth by hydrogen burning during the 
TP cycle)] at  the 8$^\mathrm{th}$  TP if different
values of $f$ are assumed. The filled squares represent 
cases where the $f$ value has been changed before the pulse-driven convective zone
appears ($t\approx -100 \mathrm{yr}$ in Fig.\,\ref{fig:KD}) while 
in the case of the open 
symbols $f$ has been changed after the pulse-driven convective zone has disappeared and before 
the onset of the third dredge-up ($t\approx +100 \mathrm{yr}$).
}
\end{figure}
While overshoot at the top boundary of the pulse-driven convective zone has no noticeable effect,
mixing beyond the bottom boundary leads to a decrease of the \ensuremath{^{4}\textrm{He}} 
abundance in the intershell. At the same time the intershell region
is enriched with \ensuremath{^{12}\textrm{C}} and \ensuremath{^{16}\textrm{O}}. This change of composition leads
to an increase of opacity and therefore the radiative gradient
\ensuremath{\nabla_\mathrm{rad}} (which is proportional to the opacity)
is lifted with respect to the adiabatic gradient \ensuremath{\nabla_\mathrm{ad}}
and the further downwards extension of the pulse-driven convective zone is supported. 
This process drives an even deeper penetration of the pulse-driven convective zone
into the core region. In Fig.\,\ref{fig:KD} the left panel shows the 
abundance profiles in the mass range of the two burning shells
shortly before the flash occurs, which is shown in the main panel.
The region between the mass coordinates $M_\mathrm{r}\approx 0.64\mathrm{M}_\odot$
and $\approx 0.65\mathrm{M}_\odot$ shows the already altered intershell
abundances (compared to standard calculations) formed during previous
thermal pulses. The almost pure He layer above $M_\mathrm{r}\approx 0.65\mathrm{M}_\odot$ 
results from H-shell burning during the previous interpulse phase.
Below $\approx 0.64\mathrm{M}_\odot$ the \ensuremath{^{4}\textrm{He}} abundance
drops continuously according to the preceding He-shell burning.
The pulse-driven convective zone of a TP computed without overshoot would not
extend into this He-poor region as can be seen from the dashed arrow in 
the main panel of Fig.\,\ref{fig:KD}. In contrast to that the pulse-driven convective zone actually
shown in the main panel is computed with overshoot and does extend much 
deeper into the core. It covers a mass  fraction which is larger by about
a factor of $1.5$ compared to the pulse-driven convective zone without overshoot. 
The similarity
of this effect with the three well known  envelope convection
dredge-up events we refer to  it as \emph{intershell dredge-up} (IS-DUP). 

The intershell abundances with overshoot is a function of time 
(TP number). While the \ensuremath{^{4}\textrm{He}}
abundance decreases steeply over the first pulses at the expense of
\ensuremath{^{12}\textrm{C}}, the \ensuremath{^{16}\textrm{O}} abundance increases steadily. After about a dozen pulses
the \ensuremath{^{4}\textrm{He}} abundance has recovered from a relative minimum (corresponding
to a relative maximum of \ensuremath{^{12}\textrm{C}}) and all three abundances level at
about (He/C/O)=(40/40/16). These
changed intershell abundances (compared to standard models) may eventually
contribute to a better understanding of the above mentioned 
abundance patterns of H-deficient post-AGB stars.

After the pulse-driven convective zone has disappeared IS-DUP has led to an intershell abundance
which contains less \ensuremath{^{4}\textrm{He}} compared to models without overshoot.
Then, following the arguments given in the previous section,
the efficiency
of the third dredge-up is enhanced. This can be seen from Fig.\,\ref{fig:f-lambda}.
If a larger overshoot efficiency is applied to the pulse-driven convective zone (filled symbols)
less \ensuremath{^{4}\textrm{He}} is present in the intershell region after the TP and 
the third dredge-up is more efficient. Thus, we find that the application
of overhoot to the pulse-driven convective zone during the He-shell flash increases the amount
of third dredge-up occuring immediately after the TP.

\begin{figure}
\plotone{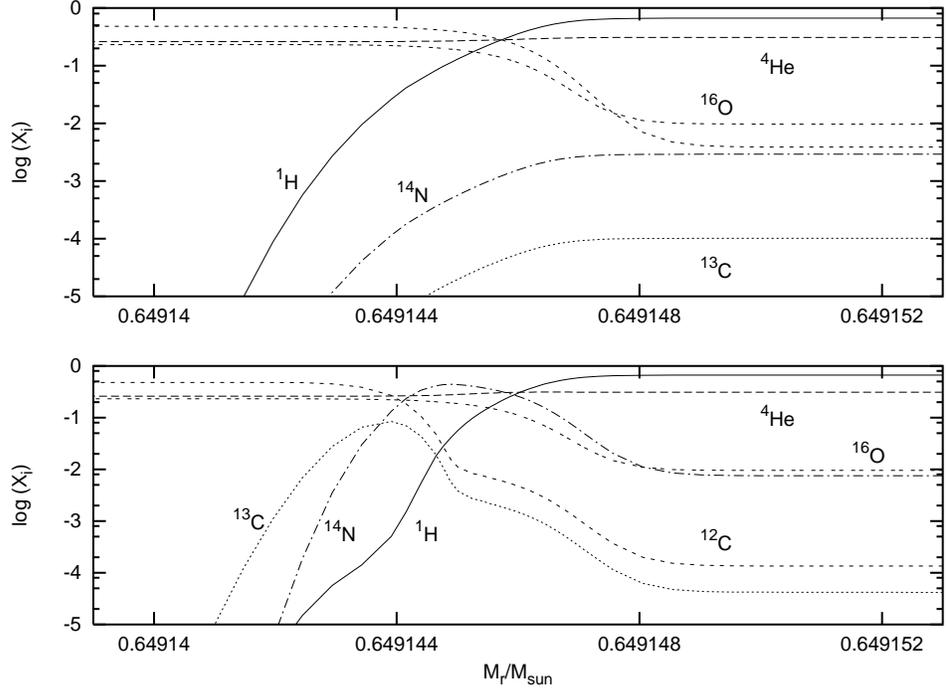}
\caption{\label{fig:C13II.ps} 
After the 8$^\mathrm{th}$ TP of a
$M_\mathrm{ZAMS}=3\mathrm{M}_\odot$ sequence:
Isotopic abundance profiles at the interface of \ensuremath{^{12}\textrm{C}}-rich intershell
(top) and H-rich envelope shortly after the end of the third dredge-up
and about $1800\mathrm{yr}$ later when the \ensuremath{^{13}\textrm{C}}-pocket has developed.
}
\end{figure}
But what is the role of overshoot at the bottom of the convective
envelope? The open symbols in Fig.\,\ref{fig:f-lambda} each represent
the dredge-up efficiency $\lambda$ when $f$ has been changed \emph{after}
the pulse-driven convective zone has disappeared. This means that all these dredge-up events
occur with identical intershell abundance (the same overshoot has
been applied to the pulse-driven convective zone). Only the efficiency 
of the overshoot at the bottom of the convective envelope is different. 
One can see that the amount of overshoot at this convective boundary
is not related to the dredge-up efficiency. The only effect of this overshoot
is the removal of the H-He discontinuity 
which would otherwise
prohibit the third dredge-up in our models ($\lambda=0$ for $f=0$ in 
Fig.\,\ref{fig:f-lambda}).
The model sequence shows the third dredge-up  with increasing efficiency 
from the third thermal pulse on
until ($\lambda > 1$). 
It leads to a transformation of the model star into a carbon star at the
13$^{th}$ TP.
We conclude that some non-zero overshoot at the
bottom of the convective envelope is necessary in our 
models to make the third 
dredge-up at all possible (However, Straniero et~al (1997) 
did find the third dredge-up for similar core masses without invocation of 
overshoot.) while overshoot at the bottom of the
pulse-driven convective zone has an effect on the efficiency of the third dredge-up.
Finally we note, that the efficient dredge-up leads to
deviations from the linear core mass - luminosity relation 
originally described 
by Paczy\'nski (1970). The deviations occur at lower
core masses than those associated with hot bottom burning.
They can be understood by the analysis of homology relations
as described by Herwig et~al (1998).

\section{The \ensuremath{^{13}\textrm{C}}-pocket}
\label{sec:cdr-pocket}
The declining mixing efficiency of our overhoot method has not been
of great importance for the effects described so far. However, 
partial mixing beyond the bottom of the envelope convection becomes
crucial when the third dredge-up comes to an end. Then the bottom
of the H-rich envelope is immediately neighbored by the \ensuremath{^{12}\textrm{C}}-rich
intershell region and due to the exponential diffusive overshoot a
thin layer forms where protons and \ensuremath{^{12}\textrm{C}} coexist (Fig.\,\ref{fig:C13II.ps}, top
panel). When the temperatures rise in this region due to
resuming contraction
after the TP a \ensuremath{^{13}\textrm{C}}-pocket (\ensuremath{^{13}\textrm{C}}-rich layer) is formed (Fig.\,\ref{fig:C13II.ps},
bottom panel). It typically contains about $3\cdot10^{-7}\mathrm{M}_\odot$
of \ensuremath{^{13}\textrm{C}}. It may, in the further course of the pulse, serve
as the neutron source for the formation of heavy elements.
While the qualitative modelling of such a pocket is quite encouraging
the \ensuremath{^{13}\textrm{C}}\ formation must be investigated more quantitatively
in the future.

\acknowledgments
  F.\ H.\ acknowledges funding by the \emph{Deut\-sche
    For\-schungs\-ge\-mein\-schaft, DFG\/} (grants Scho\,394/13
    and La\,587/16).

\end{document}